\documentclass[conference]{IEEEtran}

\usepackage{graphicx}
\usepackage{amsmath}
\usepackage{amssymb}
\usepackage[usenames]{color}
\newcommand{\diag}{ \mathrm{ diag }}

\newcommand{\vI}{ \boldsymbol{I} }

\newcommand{\vs}{ \boldsymbol{s }}
\newcommand{\vst}{ \boldsymbol{s } (t)}

\newcommand{\vAp}{ \boldsymbol{ A } (p) }

\newcommand{\vet}{ \boldsymbol{e } (t)}

\newcommand{\vyk}{ \boldsymbol{y}_k }

\newcommand{\vxk}{ \boldsymbol{x}_k }
\newcommand{\hvxk}{ \widehat{\boldsymbol{x}}_k }
\newcommand{\bvxk}{ \bar{\boldsymbol{x}}_k }

\newcommand{\vPhi}{ \boldsymbol{ \varPhi } }
\newcommand{\vPhiT}{ \boldsymbol{ \varPhi }^T }

\newcommand{\vzero}{ \boldsymbol{ 0 } }
\newcommand{\vnu}{ \boldsymbol{ \nu } }
\newcommand{\vek}{ \boldsymbol{e }_k }

\newcommand{\cN}{ \mathcal{N} }

\newcommand{\vLambda}{ \boldsymbol{ \varLambda } }
\newcommand{\vGamma}{ \boldsymbol{ \varGamma } }

\newcommand{\vRxx}{ \boldsymbol{ R }_{xx} }
\newcommand{\vRyy}{ \boldsymbol{ R }_{yy} }
\newcommand{\vRxy}{ \boldsymbol{ R }_{xy} }

\newcommand{\hs}{ \widehat{ s }}

\newcommand{\vbt}{ \boldsymbol{b} (t) }
\newcommand{\vbst}{ \boldsymbol{b}_S (t) }
\newcommand{\vbIt}{ \boldsymbol{b}_I (t) }

\newcommand{\bvA}{ \bar{\boldsymbol{A}} }

\newcommand{\bA}{ \bar{ A } }

\newcommand{\bvak}{ \bar{\boldsymbol{ a }}_k }
\newcommand{\bvaHk}{ \bar{\boldsymbol{ a }}^H_k }
\newcommand{\ktoj}{k \rightarrow j}

\newcommand{\bPsi}{ \bar{ \Psi } }

\begin{document}
\sloppy

 \title{Effectiveness of sparse Bayesian algorithm
 for MVAR coefficient estimation in MEG/EEG source-space causality analysis}

\author{\IEEEauthorblockN{Kensuke~Sekihara}
\IEEEauthorblockA{Tokyo Metropolitan University\\
Asahigaoka 6-6, Hino\\
Tokyo 191-0065, Japan\\
E-mail:kensuke.sekihara@gmail.com}
\and
\IEEEauthorblockN{Hagai Attias}
\IEEEauthorblockA{Golden Metallic Inc.\\
San Francisco, CA, USA}
\and
\IEEEauthorblockN{Julia Owen and Srikantan S. Nagarajan}
\IEEEauthorblockA{University of California, San Francisco\\
513 Parnassus Avenue \\
San Francisco, CA 94143, USA}
}

\maketitle

\begin{abstract}
This paper examines the effectiveness of a sparse Bayesian algorithm to estimate multivariate autoregressive coefficients when a large amount of background interference exists.
This paper employs computer experiments to compare two methods in the source-space causality analysis: the conventional least-squares method and a sparse Bayesian method.
Results of our computer experiments show that the
interference affects the least-squares method in a very severe manner. It produces large false-positive results, unless 
the signal-to-interference ratio is very high. 
On the other hand, the sparse Bayesian method is relatively insensitive to the existence of interference. 
However, this robustness of the sparse Bayesian method is attained on the scarifies of the detectability of true causal relationship. 
Our experiments also show that the surrogate data bootstrapping method tends to give a statistical threshold that are too low for the sparse method.
 The permutation-test-based method gives a higher (more conservative) threshold and it should be used with the sparse Bayesian method whenever the control period is available.
\end{abstract}

\section{Introduction}
Estimating a causal relationship among cortical activities using the MEG/EEG
source space analysis has gained a great interest\cite{Lin;dyn,Che;sta}.
Such
causality analysis generally requires a two step procedure: The first
step estimates source time series from MEG/EEG measurements. 
The second step computes 
some types of causality measures using the estimated time series of
target source activities. Here, popular measures are
Granger-causality-based measures.  Many
investigations have been performed in the past fifteen years to explore
the effectiveness of the Granger-causality-based measures in brain
signal analysis\cite{Gew;mea,Bac;par,Kam;ane}. 

The Granger-causality measures rely on the accurate modeling of the 
 multivariate vector auto-regressive (MVAR) process of the 
source time series.
Since, in general, the causality analysis is performed using the
source time series estimated from 
non-averaged data, 
 the estimated time series inevitably contains large
influence of brain background interference, which is often referred to
as the brain noise.  However, the MVAR modeling in general does not take such interference
into account, their existence may cause significant amount of errors in
the estimated MVAR coefficients, leading to a mis-estimation of completely wrong causality relationships. 

One approach to reduce the errors in the MVAR estimation caused due
to the background interference is to impose the sparsity constraint when
estimating the MVAR coefficients. The key assumption here is that true
brain interaction causes small number of MVAR coefficients to have
non-zero values, and most of MVAR coefficients remain to be zero. If this is
true, the sparsity constraint imposed when estimating the MVAR coefficients
should be able to prevent most MVAR coefficients to have erroneously
large values due to the influence of background interference.

This paper tests the effectiveness of a method that imposes the sparsity on its solution.
This paper uses computer experiments to compare two methods for MVAR coefficient estimation: the conventional least-squares method and a sparse Bayesian method that imposes the sparsity on its solution. In Section~\ref{method}, the two methods are briefly described. We present the results of our computer experiments to compare these two methods in Section~\ref{s_com}.  
 
\section{Method}
\label{method}
\subsection{Estimation of MVAR coefficients}
\subsubsection{Least-squares-based algorithm}
\label{LSalgo}
We assume that total $K$ multiple time series exist and we 
model these multiple time series using the 
multivariate vector 
auto-regressive (MVAR) process.
Let us denote the $k$th time series as
$s_k(t)$, where $k=1,\ldots,K$, and define a column vector $\vst$ such that 
$\vst=[s_1 (t), \ldots, s_K(t)]^T$. The time $t$ is
expressed using a unit-less value.
This $\vst$ is expressed as MVAR process, such that
\begin{equation}
\label{va}
\vst=\sum_{p=1}^P \vAp \vs (t-p) + \vet,
\end{equation}
where $\vAp$ is the $p$th coefficient matrix, $P$ is the model order, and 
$\vet$ is the residual vector.

We can estimate the MVAR coefficients
$A_{i,j}(p)$ where $i,j=1,\ldots,K$ and $p=1,\ldots,P$ based on the
least-squares principle. To derive the least-squares equation,
let us explicitly write the MVAR process for the $k$th
component $s_k (t)$ as
\begin{multline}
\label{LS-one}
s_k(t)= \sum_{j=1}^K A_{k,j}(1) s_j (t-1)+\sum_{j=1}^K A_{k,j}(2) s_j (t-2)\\
+\cdots +\sum_{j=1}^K A_{k,j}(P) s_j (t-P)+e_k (t).
\end{multline}
We assume that the source time series are obtained at
$t=1,\ldots,N_T$ where $N_T \gg K \times P$. 
Then, since Equation~(\ref{LS-one}) holds for $t=P+1,\ldots,N_T$, 
a total of $N_T -P$ linear equations are obtained by setting $t=P+1,\ldots,N_T$ in Eq.~(\ref{LS-one}). 

These equations are formulated as a matrix form,
\begin{equation}
 \label{basic_eq}
 \vyk = \vPhi \vxk + \vek.
\end{equation}
Here, the $(N_T -P) \times 1$ column vector $\vyk$ is defined as
\begin{equation}
\label{input_y}
\vyk=
\left[ 
s_k(P+1),
s_k(P+2),
\ldots, 
s_k(N_T)
\right]^T ,
\end{equation}
where the superscript $T$ indicates the matrix transpose.
 In Eq.~(\ref{basic_eq}), $\vPhi$ is an $(N_T -P) \times PK$ matrix whose $(i,j)$th element $\varPhi_{i,j}$ is given by
\begin{equation}
\varPhi_{i,j} =s_\xi \left( P-\left[ \frac{j}{K} \right] +(i-1)   \right),
\end{equation}
where
\begin{equation}
 \xi=j-\left[ \frac{j}{K} \right] K,
\end{equation}
and
$[ \cdot ]$ indicates the integer not greater than the value in parentheses.
The column vector $\vxk$ is expressed as
\begin{equation}
\label{solution_vec}
\vxk=
\left[
A_{k,1}(1),
\ldots,
A_{k,K}(1),
\ldots,
A_{k,1}(P),
\ldots,
A_{k,K}(P)
\right]^T.
\end{equation}
The residual vector $\vek$ is given by
\begin{equation}
\vek=\left[ e_k(P+1),\ldots,e_k(N_T) \right]^T.
\end{equation}
Equation~(\ref{basic_eq}) is called the Yule-Walker equation.
The least-squares estimate of $\vxk$, $\hvxk$, is then obtained using,
\begin{equation}
\label{sol_k}
 \hvxk = (\vPhiT \vPhi)^{-1}\vPhiT \vyk.
\end{equation} 

\subsubsection{Sparse Bayesian algorithm}
\label{BAalgo}
We introduce an algorithm that imposes the sparsity on its solution. 
One powerful algorithm is based on the Bayesian statistics applied to solving the linear equation in Eq.~(\ref{basic_eq}). 
In this algorithm, the prior probability distribution of
$\vxk$, $f(\vxk)$, is assumed to be Gaussian: 
\begin{equation}
 f(\vxk) = \cN(\vxk | \vzero, \vnu),
\end{equation}
where $\vzero$ is a column vector whose elements are all zero. $\vnu$
is a diagonal precision matrix. The probability distribution of
$\vyk$ given $\vxk$, $f(\vyk|\vxk)$, is also assumed to be Gaussian:
\begin{equation}
 f(\vyk|\vxk)=\cN(\vyk | \vPhi \vxk, \vLambda),
\end{equation}
where $\vLambda$ is a diagonal noise precision matrix.
Then, the posterior distribution of $\vxk$, $f(\vxk|\vyk)$, is shown to be Gaussian, and it is expressed as
\begin{equation}
f(\vxk|\vyk)=\cN(\vxk|\bvxk,\vGamma).
\end{equation}
The estimation of $\bvxk$ and $\vGamma$ is carried out by using the well-known expectation-maximization (EM) algorithm\cite{Shu;ana}. The update rules in the E~step of the EM algorithm give the estimates of those parameters, such that
\begin{alignat}{2}
 \bvxk & =\vGamma^{-1} \vPhiT \vLambda \vyk\\
 \vGamma &= \vPhiT \vLambda \vPhi + \vnu
\end{alignat}

The parameters, $\vnu$ and $\vLambda$, are estimated in the M~step of the EM algorithm. The diagonal components of $\vnu$ are obtained using
\begin{equation}
\left[ \vnu \right]_{j,j}=\frac{1}{\left[ \vRxx \right]_{j,j} }
\end{equation}
where
\begin{equation}
\vRxx=\bvxk \bvxk^T + \vGamma^{-1},
\end{equation}
and $[ \, \cdot \, ]_{j,j} $ indicates the $(j,j)$th (diagonal) element
of a matrix in parentheses.
The noise precision matrix $\vLambda$ is obtained using
\begin{equation}
\vLambda^{-1} = \diag \left[ \vRyy-\vRxy^T \vPhiT - \vPhi \vRxy +\vPhi \vRxx \vPhiT \right],
\end{equation}
where 
\begin{equation}
\vRyy=\vyk \vyk^T, \quad \mbox{and} \quad \vRxy=\bvxk \vyk^T,
\end{equation}
and $\diag[\, \cdot \,]$ indicates a diagonal matrix whose diagonal
elements are equal to those of a matrix in parentheses.
The estimate of the MVAR coefficients is given from $\bvxk$,
after the EM iteration is finished. 
This algorithm is similar to but considerably simpler than the one proposed in \cite{Pen;bay}.

\subsection{Partial directed coherence}
Once the MVAR coefficient matrices are obtained, Granger-causality-based measures can be computed. In this paper, we use the partial directed coherence (PDC) proposed in \cite{Bac;par}.
To derivce PDC, we first compute
\begin{equation}
\label{defbA}
\bvA(f) = \vI - \sum_{p=1}^P \vAp e^{-2 \pi i p f},
\end{equation}
where $\vI$ is the identity matrix with its size equal to the size of the coefficient matrix $\vAp$. Then, 
the PDC from the $k$th source to the $j$th source, $\Psi_{\ktoj}$, can be computed using
\begin{equation}
\label{def_pdc}
\Psi_{\ktoj}=\frac{|\bA_{j,k}(f)|}{\sqrt{\bvaHk  \bvak}},
\end{equation}
where $\bvak$ represents the $k$th column of the matrix $\bvA(f)$, and the superscript $H$ indicates the Hermitian transpose.

In actual measurements of brain signals, multiple trials (realizations)
are usually measured. In such cases, the above $\bvA(f)$ is computed
from each trial, and the average of each $\bvA(f)$, $\langle \bvA(f)
\rangle$, is then obtained where $\langle \cdot \rangle$ indicates the
average across trials. This $\langle \bvA(f)
\rangle$ is used to derive PDC in Eq.~(\ref{def_pdc}), such that,
\begin{equation}
\label{def_pdc2}
\Psi_{\ktoj}=\frac{|\langle \bA_{j,k}(f) \rangle|}{\sqrt{\langle \bvaHk \rangle \langle \bvak \rangle}},
\end{equation}
 In our computer
experiments, Eq.~(\ref{def_pdc2}) is used to compute PDC.

 \subsection{Statistical thresholding}
 \subsubsection{Surrogate-data bootstrapping}
\label{ss_surr}
In this paper, two types of statistical thresholding methods are tested: the surrogate-data bootstrapping\cite{The;tes} and the permutation test\cite{Goo;per}.
The surrogate-data bootstrapping is widely used in the MVAR-based causal analysis. 
In this method, the Fourier transform of the time series $s_j(t)$ is first computed; it is denoted $\sigma_j (f_k)$ where $f_k$ indicates the $k$th frequency bin. We multiply a random phase to $\sigma_j (f_k)$ to create $\sigma_j (f_k) \exp(-i \varepsilon_k)$ where $\varepsilon_k$ is a uniform random number distributed between $-\pi$ and $\pi$. The phase-modulated spectrum $\sigma_j (f_k) \exp(-i \varepsilon_k)$ is then inverse Fourier transformed to create $s_j^\ast(t)$, which is called the surrogate time series. 
This procedure is repeated for $j=1,\ldots,K$ with generating a new random number to $\varepsilon_k$.
The resultant surrogate data set $s_j^\ast(t)$ where  $j=1,\ldots,K$ does not anymore contain causal relationships, although it maintains the same power spectra as those of the original time series.

Since there are (infinitely) many ways of generating random phases, many different surrogate data sets can be generated. Let us assume that we generate total $N_B$ surrogate data sets, and compute $N_B$ different PDC results using these surrogate data sets. We denote these PDC results as $\Psi^\beta (f_k)$ where $\beta=1,\ldots,N_B$\footnote{Here, we omit the explicit notation of $\ktoj$ from the notation $\Psi^\beta (f_k)$.}.

We may derive the statistical threshold using the null distribution formed using $\Psi^\beta (f_k)$ ($\beta=1,\ldots,B$). However, such a statistical threshold does not take the multiple comparisons into account. To derive a statistical threshold that takes the multiple comparisons into consideration,  the values $\Psi^\beta (f_k)$ ($\beta=1,\ldots,B$) are standardized, such that 
\begin{equation}
 T^\beta (f_k)= \frac{\Psi^\beta (f_k)-  \bPsi^\beta (f_k) }{\sigma_B (f_k)},
\end{equation}
where $\bPsi^\beta (f_k) $ and $\sigma_B^2 (f_k)$ are the average and the variance of $\Psi^\beta (f_k)$. The maximum value of $T^\beta (f_k)$ is denoted $T_{max} (f_k)$. The null distribution for deriving the statistical threshold is formed using $T_{max} (f_k)$ from all frequency bins. That is, denoting a total number of the frequency bins $N_f$, 
 we sort $T_{max} (f_k)$ such that
\begin{equation}
 T_{max} (1) \le T_{max} (2) \le \cdots \le T_{max} (N_f).
\end{equation}
Here, $T_{max}(j)$ is the $j$th smallest value of $T_{max} (f_k)$. Setting the significance level as $\alpha$, which is the probability of occurring the type I error, the threshold value $T^{th}_{max}$ is determined as $T_{max} (\omega)$ where $\omega=(1-\alpha)N_f$. The threshold for each frequency bin $f_k$, $\Psi_{th} (f_k)$, is derived as
\begin{equation}
 \Psi_{th} (f_k)= T^{th}_{max} \sigma_B (f_k) + \bPsi^\beta (f_k) .
\end{equation}
This $\Psi_{th} (f_k)$ is the statistical threshold that takes the multiple comparisons into account. 

 \subsubsection{Permutation-test based thresholding}
 \label{ss_permu}
 In our numerical experiments, we introduce another method to derive the statistical threshold, which is based on the permutation test. To implement the permutation-test-based method, a prerequisite is that the control data\footnote{ The control data indicates the data containing only interference and sensor noise.} is available, and the method computes the MVAR coefficient also from the control data. Let us denote the MVAR coefficient computed from the $m$th-trial task data as $A^{(m)}_{i,j} (p)$ and that from the $m$th-trial control data as ${}^c \! A^{(m)}_{i,j} (p)$. Denoting the total number of trials as $N_e$, the $N_e/2$ coefficients are randomly chosen from $A^{(m)}_{i,j} (p)$ where $m=1,\ldots,N_e$ and another $N_e/2$ coefficients are randomly chosen from ${}^c\!A^{(m)}_{i,j} (p)$ where $m=1,\ldots,N_e$. These total $N_e$ coefficients are averaged, and these mean MVAR coefficients are used to compute PDC with Eq.~(\ref{def_pdc2}). Since many number of the way to choose $A^{(m)}_{i,j} (p)$ and ${}^c\!A^{(m)}_{i,j} (p)$, many number of mean MVAR coefficients can be obtained. Assuming that we have a total of $N_B$ mean MVAR coefficients, we can use $N_B$ different PDC values to form a null distribution. Therefore, using exactly the same procedure as described in Section~\ref{ss_surr}, we can derive the statistical threshold $\Psi_{th} (f_k)$.

\begin{figure}[hbt]
 \begin{center}
 \includegraphics[width=8cm]{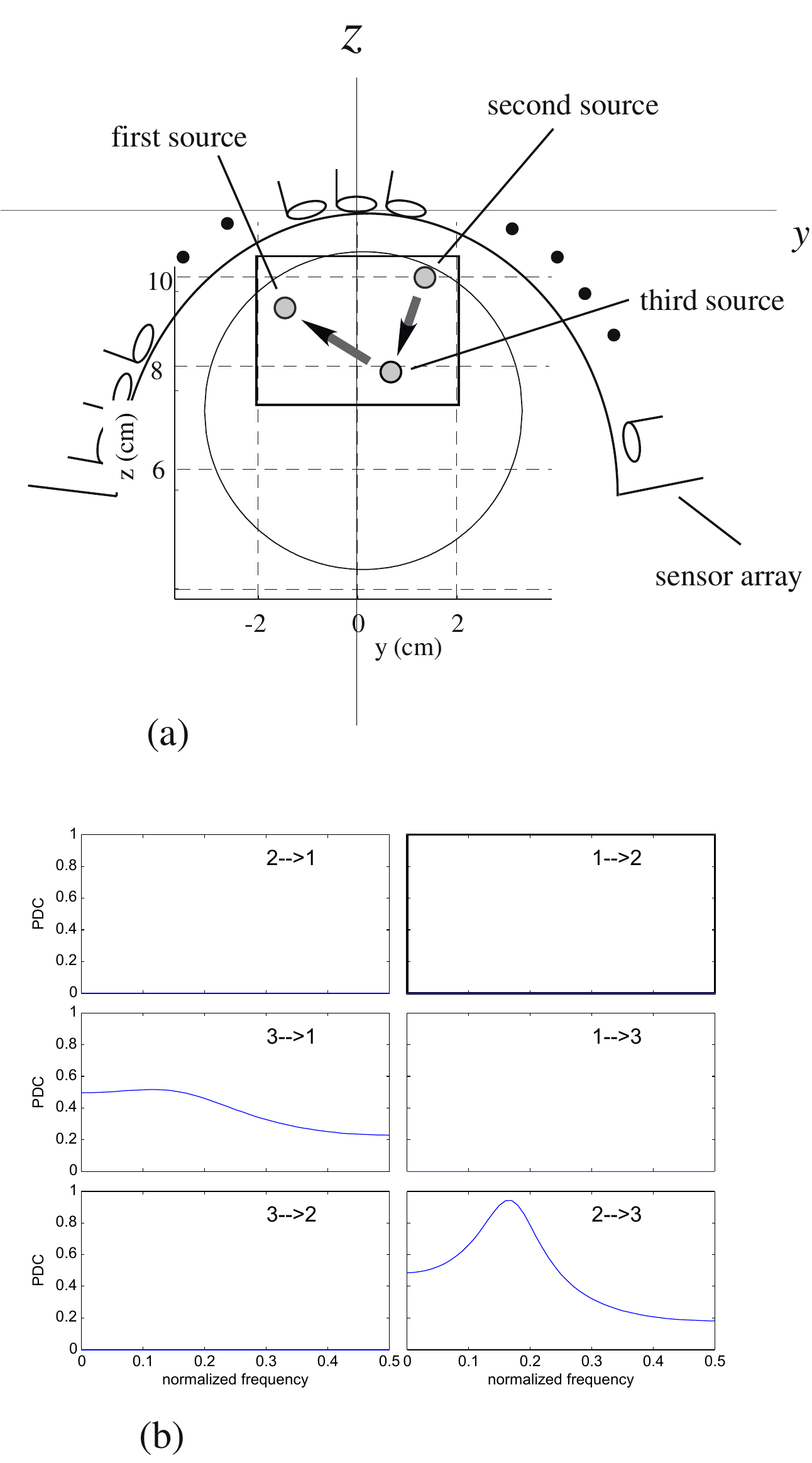}
 \caption{The coordinate system and source-sensor configuration used
	   in the numerical experiments. 
 The plane at $x=0$~cm is shown.
The small
	   circles show the locations of the three sources, and the bold arrows indicate their causal relationships assumed in the experiments. (b) The plot of partial directed coherence (PDC) computed using the model MVAR coefficients in Eq.~(\ref{genMVAR}). The ordinate of these plots indicates the frequency normalized by the sampling frequency in which the value 0.5 indicates the Nyquist frequency. The abscissa is the PDC value, which is normalized to 1.}
 \label{fig1}
 \end{center}
 \end{figure}

 \begin{figure}[hbt]
 \begin{center}
 \includegraphics[width=8cm]{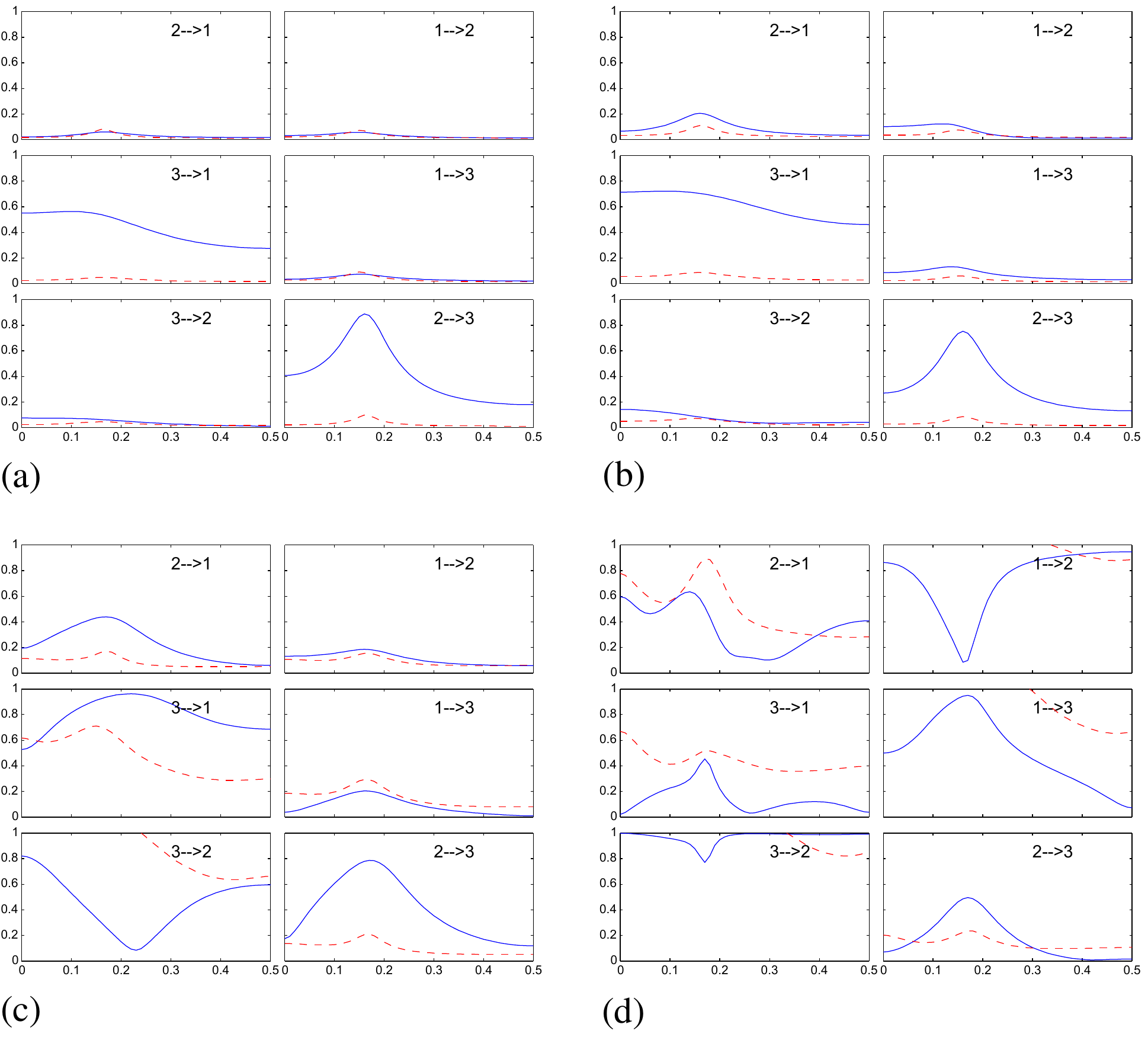}
 \caption{Results of computing partial directed coherence (PDC) for interacting source scenario. The least-squares (LS) algorithm is used. (a)Results when SIR equal to 2. (b)Results when SIR equal to 1. (c)Results when SIR equal to 0.5. (d)Results when SIR equal to 0.25.  The solid lines show PDC and the broken lines show the statistical threshold for the 95\%-significance level obtained with the surrogate-data bootstrap method.}
 \label{fig_LS}
 \end{center}
 \end{figure}
 
 \begin{figure}[hbt]
 \begin{center}
 \includegraphics[width=8cm]{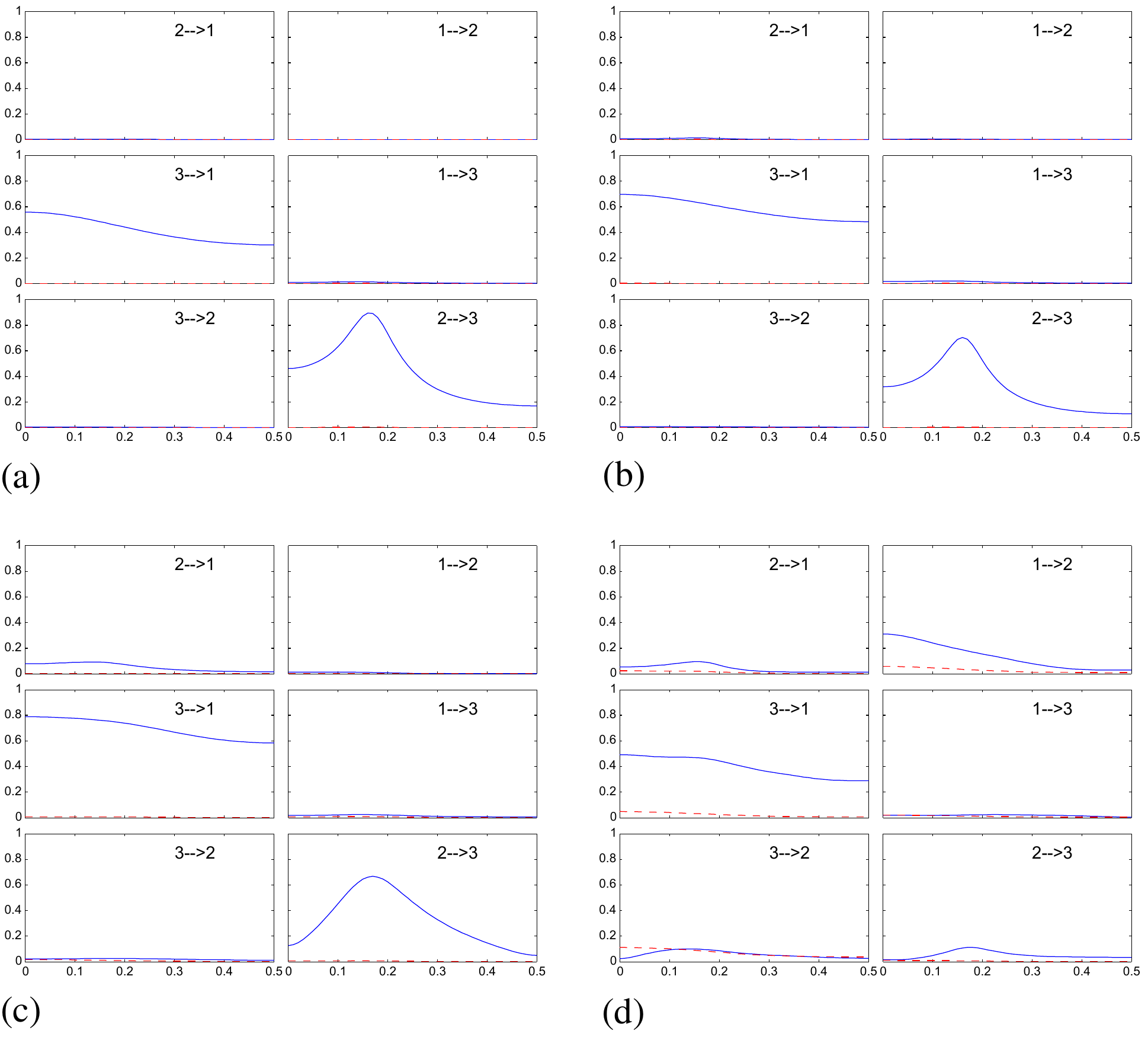}
 \caption{Results of computing partial directed coherence (PDC) for interacting source scenario. The sparse Bayesian algorithm is used. (a)Results when SIR equal to 2. (b)Results when SIR equal to 1. (c)Results when SIR equal to 0.5 (d)Results when SIR equal to 0.25.  The solid lines show PDC and the broken lines show the statistical threshold for the 95\%-significance level obtained with the surrogate-data bootstrap method. }
 \label{fig_CH}
 \end{center}
 \end{figure}

\section{Computer Experiments}
\label{s_com}
\subsection{Generation of simulated MEG recordings}
We perform numerical experiments 
to compare the performances of the algorithms described in Section~\ref{method}. In these experiments, we apply the source-space causality analysis in which source time series are first estimated from MEG recordings, and 
 the MVAR coefficients
are then computed using the source time series at selected voxels. The partial directed coherence (PDC) is computed to assess the causality relationship.  

Here, we use a sensor alignment of the 275 whole-head MEG
sensor array from Omega$^{TM}$ (VMS Medtech, Coquitlam, Canada) neuromagnetometer.
 Three sources are assumed to exist on the vertical single plane: ($x=0$~cm).
 The source-sensor configuration and 
the coordinate system are depicted in Fig.~1(a).
We assume three sources 
and the time series of these three sources are denoted $s_1(t)$,
$s_2(t)$, and $s_3(t)$.
We simulate two kinds of scenarios: interacting-source and
non-interacting-source scenarios. In the interacting source scenario,
the source activities are assumed to have causal relationships, and 
the time series of the three sources are generated using the MVAR
process reported in \cite{Din;gra}, which uses
\begin{multline}
\label{genMVAR}
\left[
\begin{array}{c}
s_1(t)\\
s_2(t)\\
s_3(t)
\end{array}
\right]
=
\left[
\begin{array}{ccc}
0.8& 0& 0.4\\0& 0.9& 0\\0& 0.5& 0.5 
\end{array}
\right]
\left[
\begin{array}{c}
s_1(t-1)\\
s_2(t-1)\\
s_3(t-1)
\end{array}
\right]\\
+
\left[
\begin{array}{ccc}
-0.5& 0& 0\\0& -0.8& 0\\0& 0& -0.2
\end{array}
\right]
\left[
\begin{array}{c}
s_1(t-2)\\
s_2(t-2)\\
s_3(t-2)
\end{array}
\right]
+
\vet.
\end{multline}
This MVAR process represents the causal relationship depicted in
Fig.~1(a), 
in which the second source has a directional causal influence 
on the third source and the third source has the directional influence on the first
source. The true PDC is computed
using the model MVAR coefficients in Eq.~(\ref{genMVAR}). The
results are shown in Fig.~1(b).
These plots of PDC is the ground truth for the following experiments.

In the non-interacting source scenario, the source activities are
assumed to 
have no causal relationships, and the
time series of the three
sources are generated as low-pass-filtered, independent random time series. 
Here,  $s_1(t)$
 is generated by applying low-pass filter with the cut-off frequency of 0.3 to a white Gaussian random time series. The time
series of the second source, $s_2(t)$, is generated by applying the
low-pass filter with the cutoff frequency of 0.2 to a white Gaussian random time series, and $s_3(t)$ is generated by applying
the low-pass filter with the cutoff equal to 0.15 to a white Gaussian random time series. (The filter cut-off frequency is expressed by the 
 the frequency normalized by the sampling frequency.) Note that, since the three white Gaussian random time series are independent, these source time series have no causal relationships.

In our computer simulation, we first generated the three-source time series, $s_1(t)$, $s_2(t)$, and $s_3(t)$, and then computed the signal magnetic recordings $\vbst$ using 
the spherical homogeneous conductor model\cite{Sar;bas}.
The simulated sensor recordings $\vbt$ were generated by
adding spontaneous MEG signal to $\vbst$, such that,
$
\vbt = \vbst + \alpha \vbIt,
$
where $\vbIt$ is the spontaneous MEG measured using the same 275 whole-head
sensor array, and $\alpha$ is a constant that controls the
signal-to-interference ratio(SIR) of the generated sensor recordings.
We generated forty-trial recordings; each consisting of 600 time
points with SIR equal to 2, 1, 0.5, and
0.25.
Note that, in general, causality analysis is performed using non-averaged trial
data, the conditions with SIR equal to 1 and 2 
should be considered high SIR conditions. The conditions with SIR below 
 0.5 are considered practical conditions for non-averaged measurements.
 
\subsection{Source-space causality analysis}
 The source
reconstruction was performed using the recently-proposed algorithm, called Champagne. 
Since the detail of this algorithm has already been published\cite{Wip;rob}, it is not described here. 
Champagne algorithm
was applied to the simulated sensor recordings $\vbt$, and 
reconstructed time series
$\hs_1(t)$, $\hs_2(t)$, and $\hs_3(t)$ were obtained as the time series at voxels
nearest to the assumed source locations.
Here, three-dimensional reconstruction was
performed on a region defined as $-4 \le x \le 4$, $-4 \le x \le 4$, and
$6 \le z \le 12$~cm with a voxel interval equal to
0.5~cm.

Once source time series are estimated, we can proceed with the MVAR
coefficient estimation using these eatimated time series,
$\hs_1(t)$, $\hs_2(t)$, and $\hs_3(t)$. 
The MVAR coefficients were estimated by using the methods
described in Section~\ref{method}. 
Here, the MVAR coefficient estimate $\hvxk$ was obtained from each
trial and, since forty trials were generated, forty sets of $\hvxk$ was
obtained. The final estimates of the MVAR coefficients
were obtained by averaging these forty sets of $\hvxk$. 
These averaged MVAR coefficients were then used to compute PDC.

\subsection{Results for interacting source scenario}
 Results for interacting source scenario
are shown in Figs.~\ref{fig_LS} and \ref{fig_CH}. 
According to the results in Fig.~\ref{fig_LS}, the least-squares method gives fairly accurate results only when
SIR is equal to 2. 
However, in other SIR cases, very
large spurious causal relationships exist in any directions. 
Although some spurious relationships
may be removed by the statistical thresholding. this is not always the case.
For example, when SIR is equal to 1 and 0.5 (Fig.~\ref{fig_LS}(b) and (c)),
the causal relationship for the ``2 to 1'' direction exceeds the
threshold, and cannot be removed by the statistical thresholding.

According to the results from the sparse algorithm in Fig.~\ref{fig_CH},
when SIR is equal to 2 or 1,
the resultant PDC is almost identical to the ideal results in Fig.~1(b).
However, when SIR is equal to 0.5, small amounts of spurious PDC arises, for example, 
in the ``2 to 1'' direction; this spurious PDC exceeds the statistical
threshold. When SIR is 0.25, spurious PDC
arises in such directions as ``2 to 1'', ``1 to 2'',  and ``3 to
2''. Also, PDC for the ``2 to 3'' direction becomes significantly
weaker than the ideal results in Fig.~1(b), indicating that the
detectability of the true causal relationship decreases due to the low SIR
condition.

\subsection{Results for non-interacting source scenario}
\begin{figure}[hbt]
 \begin{center}
 \includegraphics[width=8cm]{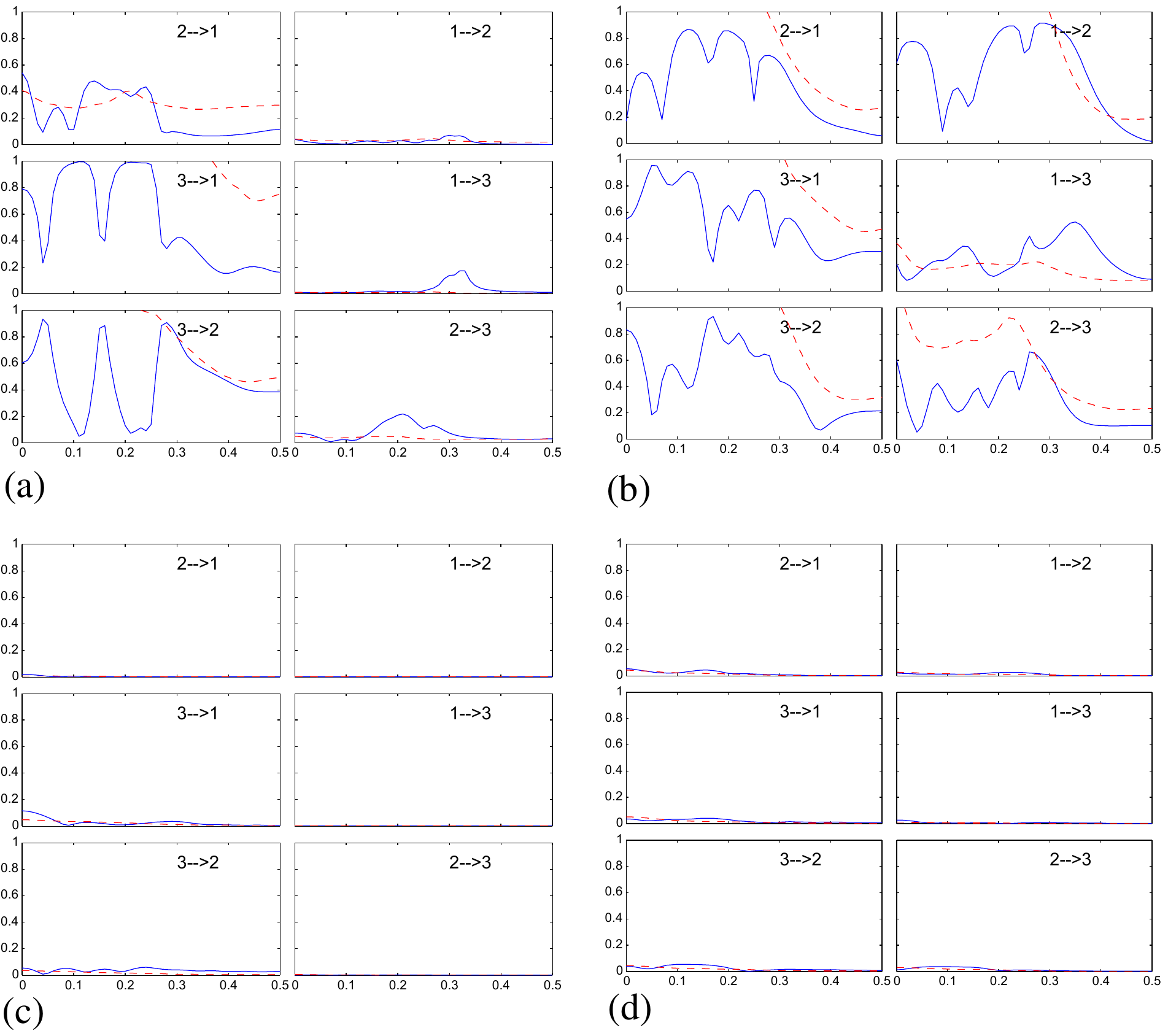}
 \caption{Results of computing partial directed coherence (PDC) for non-interacting source scenario. (a)Results from the least-squares algorithm when SIR equal to 1. (b)Results from the least-squares algorithm when SIR equal to 0.25. (c)Results from the Bayesian algorithm when SIR equal to 1. (d) Results from the Bayesian algorithm when SIR equal to 0.25.  The solid lines show PDC and the broken lines show the statistical threshold for the 95\%-significance level obtained with the surrogate-data bootstrap method.}
 \label{fig_lowpass_surro}
 \end{center}
 \end{figure}

 The results of computing PDC for non-interacting source scenario are
 shown in Fig.~\ref{fig_lowpass_surro}. 
 Since in these numerical experiments, no causal relationship is assumed
 among the activities of the three sources, the PDC for all the six
 directions must be completely zero. However, 
 the results from the least-squares method show a large amount of
 spurious PDC and the spurious PDC exceeds, in some cases, the
 statistical threshold. This happens 
 even when SIR is considerably high such as SIR equal to one. 
 On the other hand, 
 in the Bayesian results, PDC is almost completely equal to
 zero when SIR is equal to 1, but when SIR is equal to 0.25,
  a small amounts of non-zero spurious PDC exists that are sometimes exceeds the
 statistical threshold. 
 
 \subsection{Results using permutation-test based thresholding}
\begin{figure}[hbt]
  \begin{center}
  \includegraphics[width=8cm]{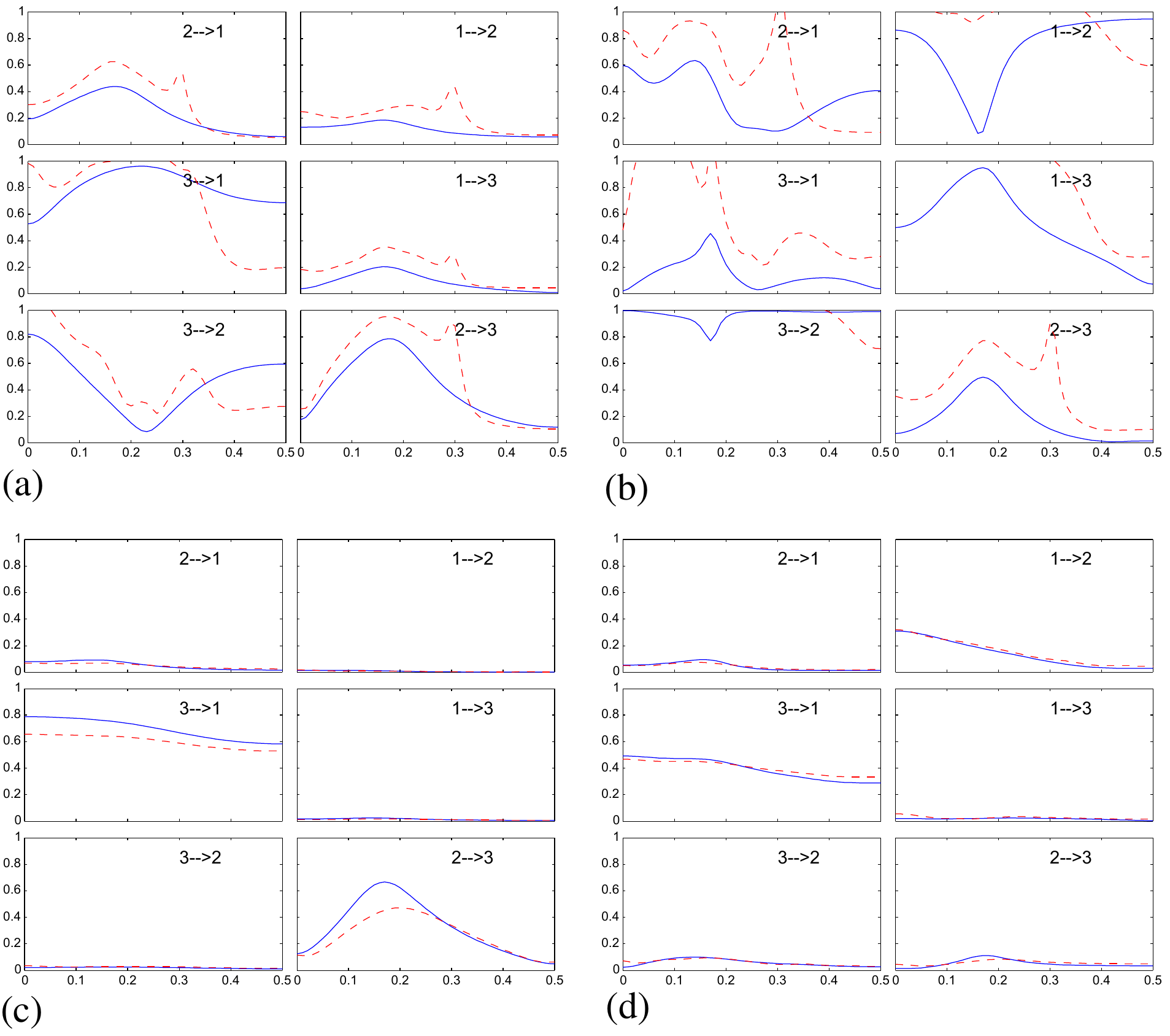}
  \caption{ Results of computing partial directed coherence (PDC) for interacting source scenario with the statistical thresholding obtained using the permutation test. (a)Results from the least-squares algorithm when SIR equal to 1. (b)Results from the least-squares algorithm when SIR equal to 0.25. (c)Results from the Bayesian algorithm when SIR equal to 1. (d) Results from the Bayesian algorithm when SIR equal to 0.25.  The solid lines show PDC and the broken lines show the statistical threshold for the 95\%-significance level obtained with the permutation-test based method.
}
  \label{fig_sig_per}
  \end{center}
  \end{figure}

\begin{figure}[hbt]
  \begin{center}
  \includegraphics[width=8cm]{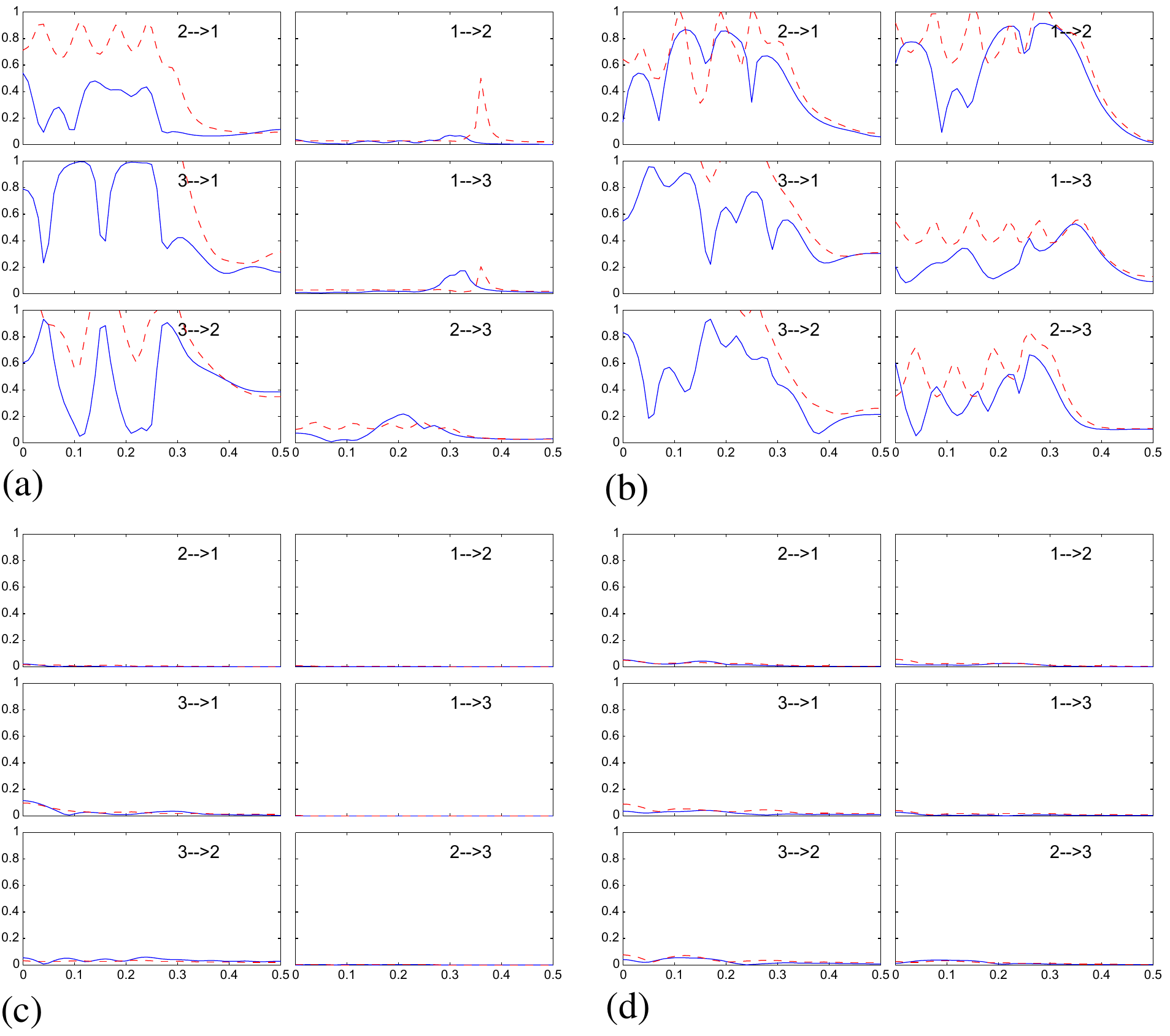}
  \caption{Results of computing partial directed coherence (PDC) for non-interacting source scenario with the statistical thresholding obtained using the permutation test. (a)Results from the least-squares algorithm when SIR equal to 1. (b)Results from the least-squares algorithm when SIR equal to 0.25. (c)Results from the Bayesian algorithm when SIR equal to 1. (d) Results from the Bayesian algorithm when SIR equal to 0.25.  The solid lines show PDC and the broken lines show the statistical threshold for the 95\%-significance level obtained with the permutation-test based method.}
  \label{fig_lowpass_per}
  \end{center}
  \end{figure}

 Next, the statistical threshold obtained using the permutation test are shown in Figs.~\ref{fig_sig_per} and \ref{fig_lowpass_per}. we can see that the statistical threshold obtained using the permutation test is more conservative than the threshold from the surrogate-data bootstrapping.
 In Fig.~\ref{fig_sig_per}(d), the threshold is as high as the spurious PDC in the ``1 to 2'' and ``2 to 1'' directions, and the spurious PDC can be thresholded out in these directions. On the other hand, the PDC in the ``3 to 1'' and ``2 to 3'' directions is also thresholded out, even though true interactions exist in these directions. 
 
 On the other hand, in Fig.~\ref{fig_lowpass_per}, the permutation-test-based method can threshold out the spurious PDC in all directions.  
 The permutation-test based thresholding can remove the spurious interactions, and it is particularly effective when used with the sparse Bayesian algorithm, since  
  it increases the protection against having false positive results, although this is attained on the sacrifices of the detectability to true interactions.

\section{Summary}
Results of our computer experiments show that the
interference affects the least-squares method in a very severe manner. It produces large false-positive results, and
 the least-squares method easily suffers from spurious
causal relationships unless 
the signal-to-interference ratio is very high, such as SIR of 2. 
On the other hand, the sparse Bayesian method is relatively insensitive to the existence of interference.
However, the robustness of the sparse Bayesian method to false positive results is attained on the scarifies of the detectability of true causal relationship. When SIR is very low,  
the sparse Bayesian method may fail to detect the true causal relationships, although
the spurious causal relationship is small.

The surrogate data bootstrapping method tends to give a statistical threshold that are too liberal for the sparse method.
This is probably because the sparsity constraint works effectively for the surrogate data, and PDC computed using the surrogate data set tends to be very small, and as a result of this, the statistical threshold becomes very low. The permutation-test-based method gives a higher (more conservative) threshold and this method is better be used with the sparse Bayesian method whenever the control period is available.

\bibliography{sekihara}

\begin{thebibliography}{10}

\bibitem{Lin;dyn}
F.-H. Lin, K.~Hara, V.~Solo, M.~Vangel, J.~W. Belliveau, S.~M. Stufflebeam, and
  M.~S. H\"am\"al\'ainen, ``Dynamic {G}ranger-{G}eweke causality modeling with
  application to interictal spike propagation,'' {\em Human Brain Mapping},
  vol.~30, pp.~1877--1886, 2009.

\bibitem{Che;sta}
B.~L.~P. Cheung, B.~Riedner, G.~Tononi, and B.~V. Veen, ``State-space
  multivariate autoregressive models for estimation of cortical connectivity
  from {EEG},'' in {\em Proceedings of 31st Annual International Conference of
  the IEEE EMBS}, (Minneapolis), pp.~61--64, September 2009.

\bibitem{Gew;mea}
J.~Geweke, ``Measures of conditional linear dependence and feedback between
  time series,'' {\em J. Am. Stat. Assoc.}, vol.~77, pp.~304--313, 1982.

\bibitem{Bac;par}
L.~A. Baccal\'a and K.~Sameshima, ``Partial directed coherence: a new concept
  in neural structure determination,'' {\em Biol. Cybern.}, vol.~84,
  pp.~463--474, 2001.

\bibitem{Kam;ane}
M.~J. Kami\'nski and K.~J. Blinowska, ``A new method of the description of the
  information flow in the brain structure,'' {\em Biol. Cybern.}, vol.~65,
  pp.~203--210, 1991.



\bibitem{Pen;bay}
W.~D. Penny and S.~J. Roberts, ``Bayesian multivariate autoregressive models
  with structured priors,'' {\em IEE Proceedings}, vol.~149, pp.~33--41, 2002.

\bibitem{Shu;ana}
R.~H. Shumway and D.~S. Stoffer, ``An approach to time series smoothing and
  forecasting using the {EM} algorithm,'' {\em Journal of Time Series
  Analysis}, vol.~3, pp.~253--263, 1982.

%
%

\bibitem{The;tes}
J.~Theiler, S.~Eubank, A.~Longtin, B.~Galdrikian, and J.~D. Farmer, ``Testing
  for nonlinearlity in time series: the method of surrogate data,'' {\em
  Physica D}, vol.~58, pp.~77--94, 1992.

%

\bibitem{Goo;per}
P.~Good, {\em Permutation tests}.
\newblock New York: Springer-Verlag, 2000.

\bibitem{Din;gra}
M.~Ding, Y.~Chen, and S.~L. Bressler, ``Granger causality: Basic theory and
  application to neuroscience,'' in {\em Handbook of Time Series Analysis}
  (B.~Schelter {\em et~al.}, eds.), pp.~500--600, Wiley-VCH, 2006.

\bibitem{Sar;bas}
J.~Sarvas, ``Basic mathematical and electromagnetic concepts of the biomagnetic
  inverse problem,'' {\em Phys. Med. Biol.}, vol.~32, pp.~11--22, 1987.

\bibitem{Wip;rob}
D.~P. Wipf, J.~P. Owen, H.~T. Attias, K.~Sekihara, and S.~S. Nagarajan,
  ``Robust bayesian estimation of the location, orientation, and time course of
  multiple correlated neural sources using {MEG},'' {\em NeuroImage}, vol.~49,
  pp.~641--655, 2010.

\end{thebibliography}
\bibliographystyle{IEEEtr}

\end{document}